\documentclass[useAMS,usenatbib]{mn2e}

\usepackage{graphicx}

\usepackage{natbib}

\usepackage{times}

\usepackage[T1]{fontenc}
\usepackage{ae,aecompl}

\newcommand{\sect}[1]{\S\,\ref{#1}}
\newcommand{\be}{\begin{displaymath}}
\newcommand{\ee}{\end{displaymath}}
\newcommand{\bea}{\begin{eqnarray}}
\newcommand{\eea}{\end{eqnarray}}

\title[Hybrid C-O-Ne progenitors of type Ia supernovae]{Hybrid C-O-Ne white dwarfs as progenitors of type Ia supernovae: dependence on Urca process and mixing assumptions}

\author[P. A. Denissenkov, J. W. Truran, F. Herwig et al.]{P. A. 
Denissenkov$^{1,4,9}$\thanks{E-mail: pavelden@uvic.ca.}, J. W. Truran$^{2,4,9}$,  
F. Herwig$^{1,4,9}$, S. Jones$^{1,5,9}$, B. Paxton$^{3}$, \newauthor
K. Nomoto$^{6,10}$, T. Suzuki$^{7}$ and H. Toki$^{8}$\\
$^{1}$Department of Physics \& Astronomy, University of Victoria,
       P.O.~Box 1700, STN CSC, Victoria, B.C., V8W~2Y2, Canada\\
$^{2}$Department of Astronomy and Astrophysics, and Enrico Fermi Institute, University of Chicago, Chicago, IL 60637 USA\\
$^{3}$Kavli Institute for Theoretical Physics and Department of Physics, Kohn Hall, University of California, Santa Barbara, CA 93106, USA\\
$^{4}$The Joint Institute for Nuclear Astrophysics, Notre Dame, IN 46556, USA\\
$^{5}$Astrophysics Group, Research Institute for the Environment, Physical Sciences and Applied Mathematics, Keele University, Keele, \\
Staffordshire ST5 5BG\\
$^{6}$Kavli Institute for Physics and Mathematics of the Universe (WPI), The University of Tokyo, Kashiwa, Chiba 277-8583, Japan\\
$^{7}$Department of Physics, College of Humanities and Sciences, Nihon University, Sakurajosui 3-25-40, Setagaya-ku, Tokyo 156-8550, Japan\\
$^{8}$Research Center for Nuclear Physics, (RCNP), Osaka University, Ibaraki, Osaka 567-0047, Japan\\
$^{9}$NuGrid collaboration\\
$^{10}$Hamamatsu Professor}
\begin{document}

\date{Accepted 2014 December 31. Received 2014 December 31; in original form 2014 December 31}

\pagerange{\pageref{firstpage}--\pageref{lastpage}} \pubyear{2014}

\maketitle

\label{firstpage}

\begin{abstract}
When carbon is ignited off-centre in a CO core of a super-AGB star, its burning
in a convective shell tends to propagate to the centre.
Whether the C flame will actually be able to reach the centre depends on the efficiency
of extra mixing beneath the C convective shell. Whereas
thermohaline mixing is too inefficient to interfere with the C-flame
propagation, convective boundary mixing can prevent the C burning from reaching
the centre. As a result, a C-O-Ne white dwarf (WD) is formed, after the star has
lost its envelope. Such a ``hybrid'' WD has a small CO core surrounded by a thick
ONe zone. In our 1D stellar evolution computations,
the hybrid WD is allowed to accrete C-rich material, as if it were in a close binary system
and accreted H-rich material from its companion with a sufficiently high rate at which
the accreted H would be processed into He under stationary
conditions, assuming that He could then be transformed into C. When the mass of the
accreting WD approaches the Chandrasekhar limit, we find a series of convective
Urca shell flashes associated with high abundances of $^{23}$Na and $^{25}$Mg.
They are followed by off-centre C ignition leading to convection that occupies
almost the entire star. To model the Urca processes, we use the most recent well-resolved data 
for their reaction and neutrino-energy loss rates.
Because of the emphasized uncertainty of the convective Urca process in our hybrid WD models of SN Ia progenitors, we
consider a number of their potentially possible alternative instances for different mixing assumptions, all of which
reach a phase of explosive C ignition, either off or in the centre.
Our hybrid SN Ia progenitor models have much lower C to O abundance ratios at the moment of the explosive C
ignition than their pure CO counterparts, which may explain the observed diversity of the SNe Ia.
\end{abstract}


\section{Introduction}
\label{sec:intro}

Type Ia supernovae (hereafter, SNe Ia) are thermonuclear explosions of carbon-oxygen white dwarfs (CO WDs) \citep{hillebrandt:00}.
When the mass of a CO WD approaches the Chandrasekhar limit $M_\mathrm{Ch}\approx 1.39\,M_\odot$ and, as a consequence,
the star is about to lose its hydrostatic equilibrium, C burning near the centre enters 
the phase of a thermonuclear runaway leading to an SN Ia explosion. SNe Ia are important objects because
they are the main producers of iron peak elements and also because of their single-parameter
light curves \citep{phillips:93} that allow to measure their peak luminosities with a high precision. The latter property of
SNe Ia has made them a standard candle in extragalactic astronomy, which was used to discover the accelerating
expansion of the Universe \citep{riess:98,perlmutter:99}.

CO WDs are the cores of asymptotic giant branch (AGB) stars that get exposed after their parent stars have lost their envelopes
via stellar winds or unstable mass transfers in binary systems. Masses of newly born CO WDs range from $\sim 0.6\,M_\odot$
to $\sim 1\,M_\odot$, therefore the CO WDs need to accrete some extra mass for them to reach $M_\mathrm{Ch}$.
For SNe Ia, such accretion has been proposed to occur either through a single degenerate (SD) channel, when
the extra mass is provided by WD's main sequence, subgiant, or red giant companion, or through a double
degenerate (DD) channel, when two CO WDs with a total mass larger than $M_\mathrm{Ch}$ merge \citep{wang:12}. 
There also exists a possibility that a CO WD with a mass less than $M_\mathrm{Ch}$
accretes He from its binary companion, which became a He star earlier, until the accreted layer of 
$\sim 0.04$\,--\,$0.08\,M_\odot$ of He detonates near its base causing a detonation of 
the underlying CO WD \citep{nomoto:82b,iben:91,woosley:94,shen:09}.

There is an upper limit $M_\mathrm{u}$ for the initial mass of a star $M_\mathrm{i}$, such that in a star
with $M_\mathrm{i} > M_\mathrm{u}$ carbon is ignited off-centre in the CO core on the AGB. Such
stars are called super AGB stars. If no other mixing is present
in the CO core, except convection driven by the C burning, then the C flame propagates all the way down to the centre
resulting in the formation of an oxygen-neon core that will become an ONe WD after the envelope is lost. Therefore,
only the stars with $M_\mathrm{i} < M_\mathrm{u}$ are usually considered to be able to leave remnants that may lead 
to SN Ia explosions, provided that they initially reside in binary systems with parameters suitable for the SD or DD channel.
A theoretical value of $M_\mathrm{u}$ depends on various model parameters and assumptions, such as the metallicity,
or the heavy-element mass fraction $Z$ \citep{becker:79}, the adopted carbon burning rate, that is still very uncertain 
\citep{chen:14}, as well as on the extent of convective boundary mixing outside the H and He convective cores and 
stellar rotation, both of which work towards increasing the final mass of the H-free core, hence decreasing $M_\mathrm{u}$.

There are several potential mixing mechanisms that can operate below the C-burning convective shell in the CO core.
Two of them, thermohaline mixing and convective boundary mixing (CBM), have been studied by \cite{denissenkov:13} using
the most recent estimates of their efficiencies in stars obtained from multi-dimensional numerical simulations.
It has been shown that thermohaline mixing cannot interfere with the C-flame propagation, in spite of quite a strong
positive mean molecular weight gradient produced by the off-centre C burning, because of a small aspect ratio of
participating fluid parcels (``salt fingers''). On the other hand, even a small amount of CBM turns out to be sufficient to remove
physical conditions required for the C-flame to propagate all the way to the centre. As a result, the C flame stalls
at some distance from the centre, leaving below a small core of unburnt C and O. The works of \cite{denissenkov:13} and
\cite{chen:14} have introduced a new class of
hybrid cores of super-AGB stars and their corresponding WDs that consist of small CO cores surrounded by thick ONe zones.

In this paper, we aim to find out if the hybrid C-O-Ne WDs can be progenitors of SNe Ia. In the case of a positive answer,
this will slightly increase the upper limit for the initial mass of the star that can potentially end up as an SN Ia,
from $M_\mathrm{u}$ to $M_\mathrm{u} + \Delta M_\mathrm{u}^\mathrm{hy}$, assuming that stars with
$M_\mathrm{i} < M_\mathrm{u}$ produce pure CO WDs, while those with
$M_\mathrm{i} > M_\mathrm{u} + \Delta M_\mathrm{u}^\mathrm{hy}$ give birth to pure ONe WDs.
What is probably even more important is that the hybrid C-O-Ne WDs may become progenitors of unusual SNe Ia, 
thus contributing to their diversity, because chemical and thermal
structures of the hybrid SN Ia progenitors are expected to be very different from those of the standard CO progenitors. 
In particular, we anticipate that
Urca processes will be essential for the hybrid SN Ia progenitors, because C burning usually results in a relatively
high abundance of $^{23}$Na that forms an Urca pair with $^{23}$Ne.

The structure of our paper is simple. In \sect{sec:method}, we describe how we calculate our models, including
our treatment of the Urca processes. Our results are presented in \sect{sec:results}, which
is followed by the final section where we discuss the results and make conclusions.

\section{Computational Method}
\label{sec:method}

In this section, we describe the computer code and methods that we have used to calculate our SN Ia progenitor models and we
discuss our treatment of the mass accretion and Urca processes.

\subsection{Computer code}

We have used the revision 5329 of MESA (the Modules for Experiments in Stellar Astrophysics) to prepare models of CO, ONe and
hybrid C-O-Ne WDs and to compute their evolution, driven by accretion, towards the beginning of SN Ia explosion or 
electron-capture induced collapse.
MESA\footnote{{\tt http://mesa.sourceforge.net}} is a suite of an open source modern
stellar evolution code and related modules \citep{paxton:11,paxton:13}.
We employ the same MESA equation of state, opacities and reaction rates as in the work of \cite{denissenkov:13}.
Our nuclear network is similar to the largest one ({\tt nova.net}) with 77 isotopes from H to 
$^{40}$Ca coupled by 442 reactions that was used by \cite{denissenkov:14} to prepare WD models for their simulations of
ONe nova outbursts, except that it has been extended to 87 isotopes and 500 reactions
to include the MESA electron capture network and a new Urca network that become important at very high densities.
The Urca network makes use of the well resolved tables of reaction and neutrino-loss rates for Urca pairs with 
atomic numbers $A=23$, 25 and 27 calculated by \cite{toki:13}. A detailed discussion of the new Urca rates and their comparison with
the old data of \cite{oda:94} can be found in the paper of \cite{jones:13}.

Our stellar evolution models have $Z=0.014$ and the initial hydrogen mass fraction $X=0.70$. In convective zones, 
we use the mixing length theory (MLT) of Cox \& Giuli \citep{weiss:04} with a solar calibrated ratio 
$\alpha_\mathrm{MLT} = \Lambda/\lambda_P = 1.91$ of the mixing length to the pressure scale height. 
The CBM outside a convective zone is modeled with the diffusion coefficient
\bea
D_\mathrm{CBM}(r) = D_\mathrm{MLT}(r_0)\exp\left(-\frac{|r-r_0|}{f\lambda_P}\right),
\label{eq:DCBM}
\eea
where $D_\mathrm{MLT}(r_0) = (\Lambda v_\mathrm{conv})/3 $ is a value of the convective diffusion coefficient
at a distance $f\lambda_P$ from the Schwarzschild boundary inside the convective zone calculated using local MLT values of
$\Lambda$ and convective velocity. Like \cite{denissenkov:13} and \cite{chen:14}, we use the CBM parameter $f=0.014$ 
for the H and He convective core boundaries, while we assume $f=0.007$ for the stiffer boundaries of the He and C convective shells.

\begin{figure}
\includegraphics[width=9cm]{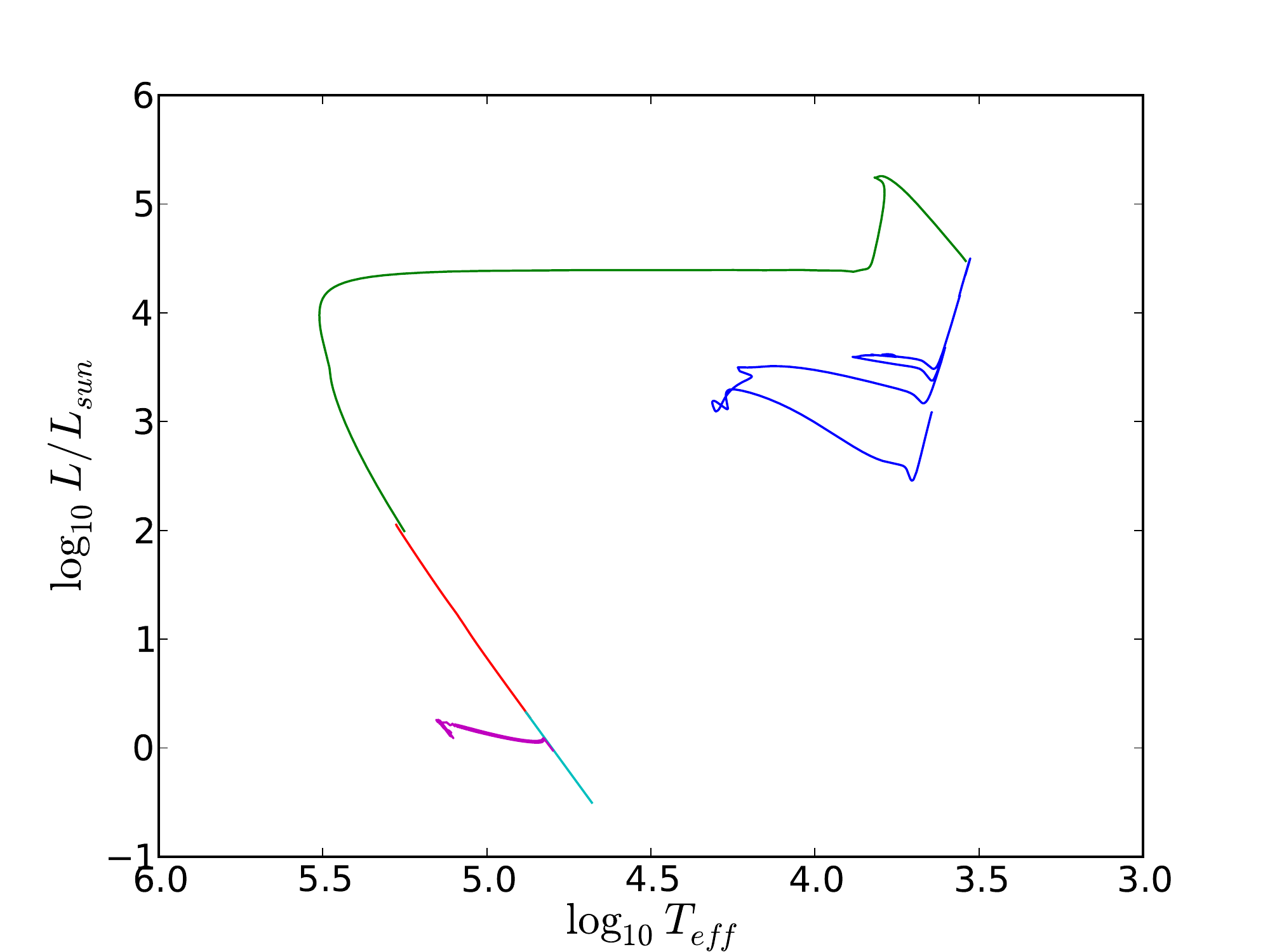}
\caption{\small Evolutionary track of the star with $M_\mathrm{i} = 6.3\,M_\odot$ calculated from the pre-main sequence
         through to the WD cooling sequence (the dark blue, green, red and light blue segments)
         using the MESA test suite case {\tt make\_co\_wd}. After that, the WD is allowed to accrete C-rich material
         until its mass approaches the Chandrasekhar limit and, as a consequence, C is ignited in the centre
         and then its burning enters the phase of thermonuclear runaway (the magenta segment).
         }
\label{fig:hrd_m63}
\end{figure}

To create a set of WDs with different initial masses and chemical structures, we use the new MESA test suite case {\tt make\_co\_wd}.
We calculate the evolution of a star from the pre-main sequence to the end of core C burning in super-AGB stars or
to the first He-shell thermal pulse in AGB stars that do not ignite C (the dark blue
segment of the track in Fig.~\ref{fig:hrd_m63}). Then, we 
artificially increase the mass-loss rate (the green segment), while using the new MLT++ prescription for
radiation-dominated envelopes with super-adiabatic convection to keep calculation timesteps reasonably large
\citep{paxton:13}. After that, we allow our model to lose the rest of its envelope until neither H nor He            
is left at its surface (the red segment). The naked WD is then cooled down to a specified central temperature $T_\mathrm{c}$
(the light blue segment). The resulting mass fraction profiles of $^{12}$C in the cores of AGB and super-AGB stars immediately
before the mass-loss rate is increased are shown in Fig.~\ref{fig:hy_wd_c12}. These include pure CO and hybrid C-O-Ne cores.

\subsection{Mass accretion}

In the SD channel, a WD accretes H-rich material from its binary companion. 
There is a lower limit $\dot{M}_\mathrm{stable}$ for the mass accretion rate 
$\dot{M}_\mathrm{acc}$, such that only for $\dot{M}_\mathrm{acc} > \dot{M}_\mathrm{stable}$ the accreted H is transformed
into He in a stable nuclear burning \citep{nomoto:82a}. 
Lower accretion rates lead to H shell flashes followed by ejecta of processed material,
resulting in events known as nova outbursts \citep[e.g.,][]{denissenkov:14}. Spectroscopic observations of ejecta of classical novae
reveal that most of them are strongly enriched in heavy elements, which is interpreted as a result of mixing with underlying
WDs \citep[e.g.,][]{gehrz:98}. Therefore, masses of WDs in the classical nova systems are unlikely to increase with time.
However, this is probably not true for the recurrent nova systems, in which WDs with masses close to the Chandrasekhar limit
accrete H with rates close to $\dot{M}_\mathrm{stable}$. This makes them good candidates for the SD channel of SN Ia progenitors
\citep{wolf:13,tang:14}. 

There is also an upper limit $\dot{M}_\mathrm{RG}$, such that H accreted by a WD with a rate
$\dot{M}_\mathrm{acc} > \dot{M}_\mathrm{RG}$ cannot be transformed into He fast enough to avoid its accumulation around the WD.
This leads to expansion of the accreted H-rich envelope, so that the accreting star tends to become a red giant.
To expand the narrow range of $\dot{M}_\mathrm{acc}$ suitable for the SD channel, \cite{hachisu:96} and \cite{ma:13}
proposed to replace the red-giant regime with the thick-wind and super-Eddington regimes, respectively, in either of which
the excess of mass that is accreted with $\dot{M}_\mathrm{acc} > \dot{M}_\mathrm{RG}$ but not processed in the H-shell burning 
is blown off the binary system.

\begin{figure}
\includegraphics[width=9cm]{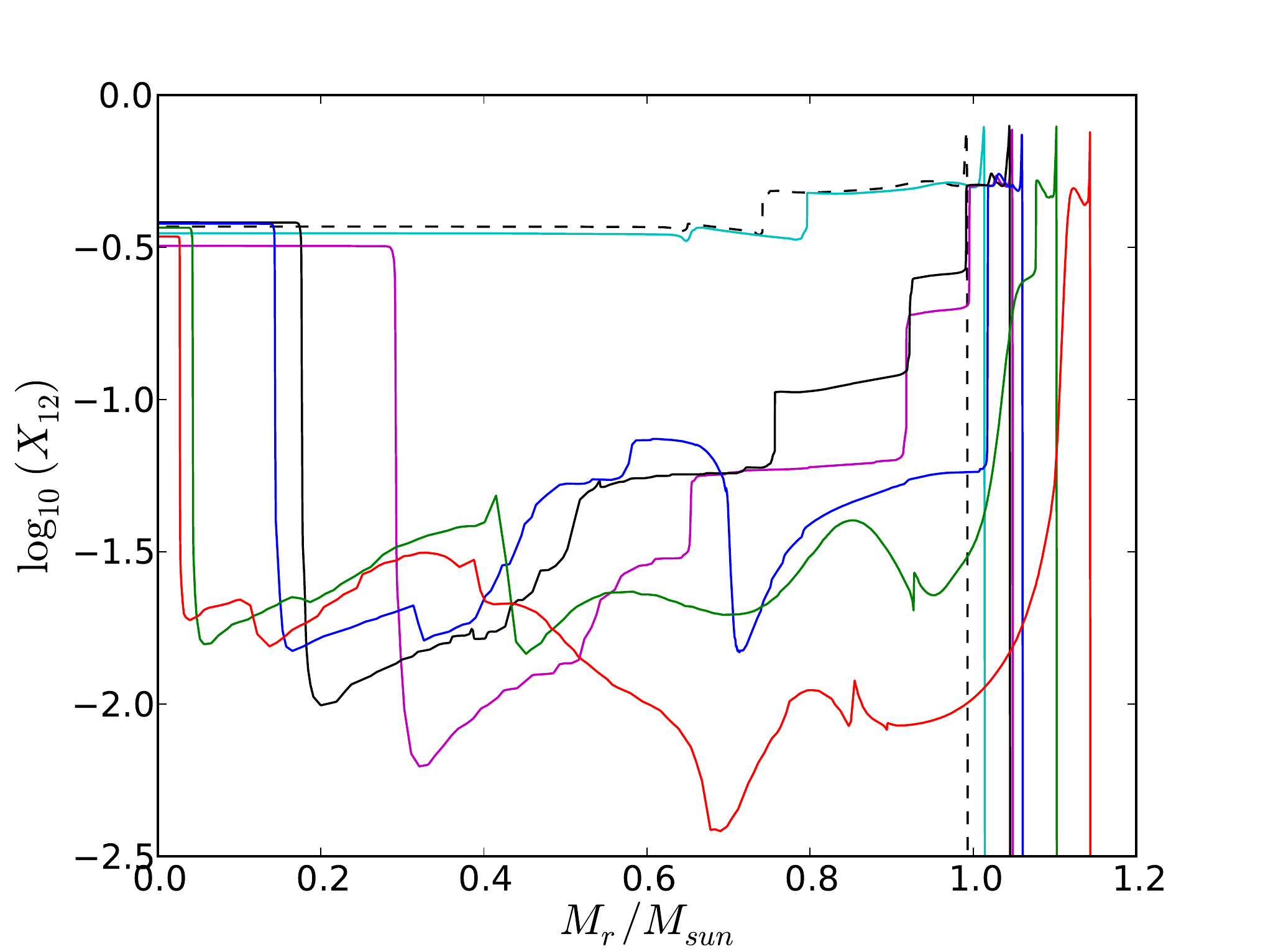}
\caption{\small Mass fraction profiles of $^{12}$C at the end of carbon burning in hybrid C-O-Ne cores of
         super-AGB stars. A value of $f = 0.014$ was used to model convective boundary mixing (CBM) at all convective
         boundaries, except the stiffer boundaries of the He and C convective shells, where a value of $f = 0.007$
         was assumed. The initial masses of the AGB stars are $6.3\,M_\odot$ (dashed black curve),
         $6.4\,M_\odot$ (cyan curve), $6.5\,M_\odot$ (magenta curve), $6.8\,M_\odot$ (solid black curve),
         $6.9\,M_\odot$ (blue curve), $7.1\,M_\odot$ (green curve), and $7.3\,M_\odot$ (red curve).
         In the first two models, C has not ignited in the CO cores.
         }
\label{fig:hy_wd_c12}
\end{figure}

To find out if the hybrid C-O-Ne WDs can be progenitors of SNe Ia, we allow our hybrid WD models to accrete material with
chemical composition identical to that of their surface layers. Because our hybrid WD models have CO-rich surface buffer zones
(Fig.~\ref{fig:hy_wd_c12}), their surface chemical composition corresponds to that of products of the He shell burning,
and so does the composition of the accreted material. In this case, we do not have problems with
the H burning and He shell flashes in the accreted envelope, like those encountered by \cite{cassisi:98}. 
Therefore, we use accretion rates that are not necessarily always in the range
$\dot{M}_\mathrm{stable} < \dot{M}_\mathrm{acc} <  \dot{M}_\mathrm{RG}$,
but with which we can relatively quickly calculate the evolution of our accreting WDs till the beginning of
an SN Ia explosion. With this simple method, we can neglect all the uncertainties associated with the accretion of H-rich
material onto a WD in the SD channel, such as the H and He retention efficiencies \citep{bours:13}, while trying to find out
if C in our hybrid WD models will eventually ignite and its burning enter the phase of thermonuclear runaway when their masses 
approach the Chandrasekhar limit. By the same reason, we do not make an evolutionary motivated choice of
the initial central temperatures of our WDs when the accretion begins. 

For a more systematic study 
of the evolution of pure CO WDs from the onset of accretion to the explosive C ignition
for a set of $T_\mathrm{c}$ and $\dot{M}_\mathrm{acc}$ values suitable for the SD channel, the interested reader
is referred to the recent study by \cite{han:14}, who also used MESA and assumed the accretion of CO-rich material,
although they did not include Urca reactions.

\begin{figure}
\includegraphics[width=9cm]{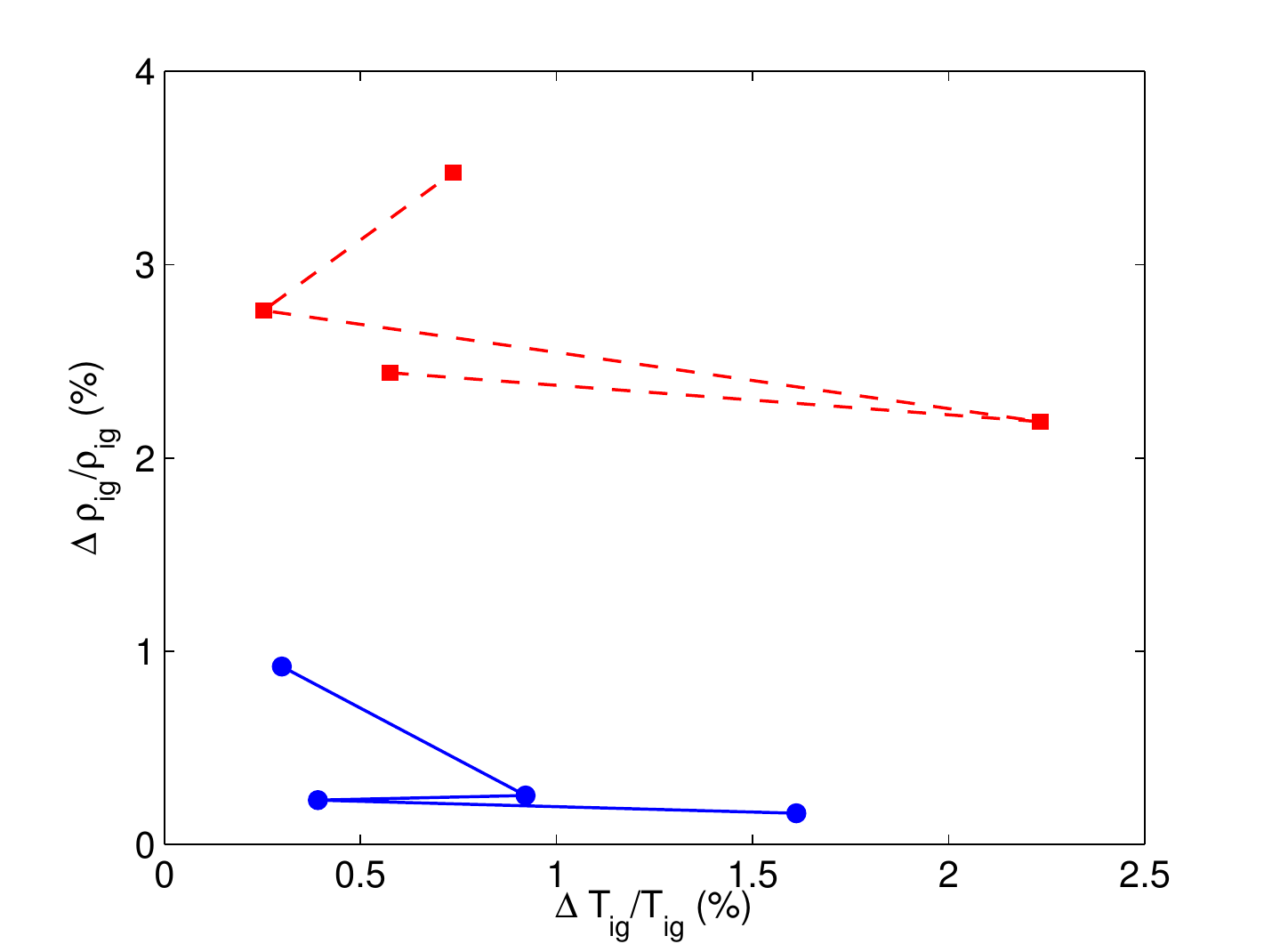}
\caption{\small Relative differences in temperature and density at the explosive C ignition between two sets of
         CO WD models accreting CO-rich material, according to data from Tables 1 and 2 of \protect\cite{han:14}. 
         The first set, like our models, neglects both H and He burning, whereas in the second set their energy generation rates
         are approximated by an extra source of $10^5 \mbox{erg g}^{-1}\mbox{s}^{-1}$ included in the outermost zones of 
         the accreting WD models. The data are plotted for the $1.0 M_\odot$ WD model with the cooling times 
         (when the accretion begins) ranging from 0.01 to 0.2 Gyr (solid blue curves corresponding to central C ignition) 
         and from 0.4 to 1.5 Gyr (dashed red curves for off-centre C ignition).
         }
\label{fig:chen_rhoT}
\end{figure}

In our considered cases, the luminosity of an accreting WD is actually determined by a release of the gravitational 
energy of accreted material as well as by the stationary H burning alternating with He shell flashes. 
Our neglect of the H and He thermonuclear energy sources results in a relatively low luminosity of our accreting WD model 
(e.g., Fig.~\ref{fig:hrd_m63}). To estimate possible effects of this neglect on the SNIa progenitor models, 
\cite{han:14} have included an extra source with the energy generation rate of
$10^5 \mbox{erg g}^{-1}\mbox{s}^{-1}$ in the outermost zones of their WD models that, like our models, accrete CO-rich material.
Fig.~\ref{fig:chen_rhoT}, in which we have used data from Tables 1 and 2 of the cited paper, shows that the relative differences in 
temperature and density at the explosive C ignition between the two sets of models, the one with and the other without the extra
source simulating the energy generation in the H and He burning, do not exceed a few percent even for WD models with very long
cooling times in which C is ignited off-centre.

For the accreted material, the MESA stellar evolution code uses the thin-shell radiation calculation of the gravitational energy
generation rate \citep[Equation 18 in][]{paxton:13}. It is obtained under the realistic assumption that the thermal
timescale is much shorter than the local accretion time in the outermost layers, therefore the accreted material adjusts
its temperature to a value nedded to transport the stellar luminosity from the underlying layers out.

\subsection{The Urca process uncertainty}

\begin{figure}
\includegraphics[width=9cm]{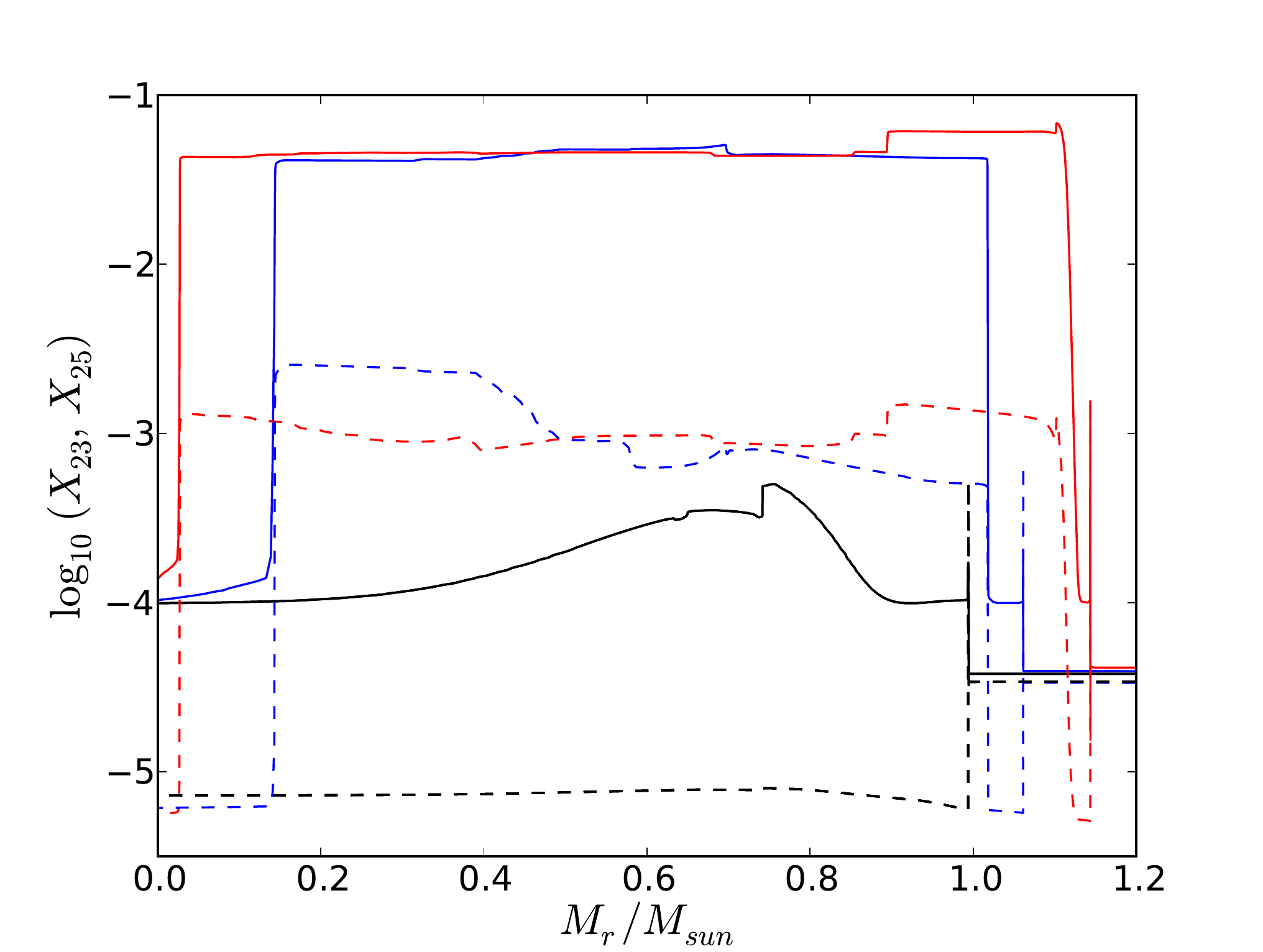}
\caption{\small Mass fraction profiles of $^{23}$Na (solid curves) and $^{25}$Mg (dashed curves) 
         in the CO core of the AGB star with $M_\mathrm{i} = 6.3\,M_\odot$ (black curves) after the first He-shell thermal pulse, 
         and in the hybrid C-O-Ne cores of the super-AGB stars with the initial masses $6.9\,M_\odot$ 
         (blue curves) and $7.3\,M_\odot$ (red curves) at the end of C burning.
         }
\label{fig:x23x25}
\end{figure}

The Urca process is an electron capture reaction that transforms a mother nucleus M into a daughter nucleus D followed by
a beta decay of D back to M. During this process, a neutrino and an anti-neutrino are emitted that escape
the star carrying away some energy, which leads to a local cooling \citep{gamow:41}. The Urca process starts when 
the electron chemical potential $\mu_\mathrm{e}$ exceeds a threshold value $\mu_\mathrm{th}$ for the corresponding Urca pair M/D.
In the electron degenerate matter, $\mu_\mathrm{e}$ mainly depends on the density $\rho$, therefore the condition
for the occurrence of the Urca process is $\rho > \rho_\mathrm{th}$. In compact stars with high central densities $\rho_\mathrm{c}$,
like the WDs with masses approaching the Chandrasekhar limit, only those Urca pairs are important that have
relatively high abundances of M nuclei and for which $\rho_\mathrm{c} > \rho_\mathrm{th}$. In our case, these are the pairs
$^{25}$Mg/$^{25}$Na with $\rho_\mathrm{th} = 1.29\times 10^9$ g\,cm$^{-3}$ ($\log_{10}\rho_\mathrm{th} = 9.11$) and,
especially, $^{23}$Na/$^{23}$Ne with $\rho_\mathrm{th} = 1.66\times 10^9$ g\,cm$^{-3}$ 
($\log_{10}\rho_\mathrm{th} = 9.22$)\footnote{The density estimates are obtained for the electron number fraction 
$Y_\mathrm{e} = 0.5$}. The mass fraction profiles of $^{23}$Na and $^{25}$Mg in the cores of three stars
from Fig.~\ref{fig:hy_wd_c12} are shown in Fig.~\ref{fig:x23x25}. The abundances of $^{23}$Na
are much larger in the super-AGB stars because of C burning in the reaction 
$^{12}$C$\,+^{12}$C$\, \rightarrow ^{23}$Na$\,+^{1}$H.

Using the electron-capture rate $\lambda^+$, beta-decay rate $\lambda^-$ and their corresponding neutrino-energy loss rates
$L^\pm$ per one M and D nucleus from the tables of \cite{toki:13}, we can calculate a total rate of change of thermal energy
in the so-called ``Urca shell'', where $\rho = \rho_\mathrm{th}$. Per unit volume, it is equal to
\bea
\varepsilon_\mathrm{U} = -\delta\mu\,\frac{d\,n_\mathrm{M}}{dt} - L^+n_\mathrm{M} - L^-n_\mathrm{D},
\label{eq:epsU}
\eea
where $n_\mathrm{M}$ and $n_\mathrm{D}$ are number densities of the M and D nuclei,
\bea
\frac{d\,n_\mathrm{M}}{dt} = -\lambda^+n_\mathrm{M} + \lambda^-n_\mathrm{D},
\eea
and
\bea
\delta\mu = \mu_\mathrm{e} - \mu_\mathrm{th} = (\mu_\mathrm{e} - V_\mathrm{s}) - (Q-\delta Q) - m_\mathrm{e}c^2.
\eea
The screening corrections to the electron potential and reaction $Q$-value, $V_\mathrm{s}$ and $\delta Q$, used in
the last equation are also provided by the tables. Note that $\delta Q$ is negative 
for beta decays and positive for e-captures.

\begin{figure}
\includegraphics[width=8.5cm]{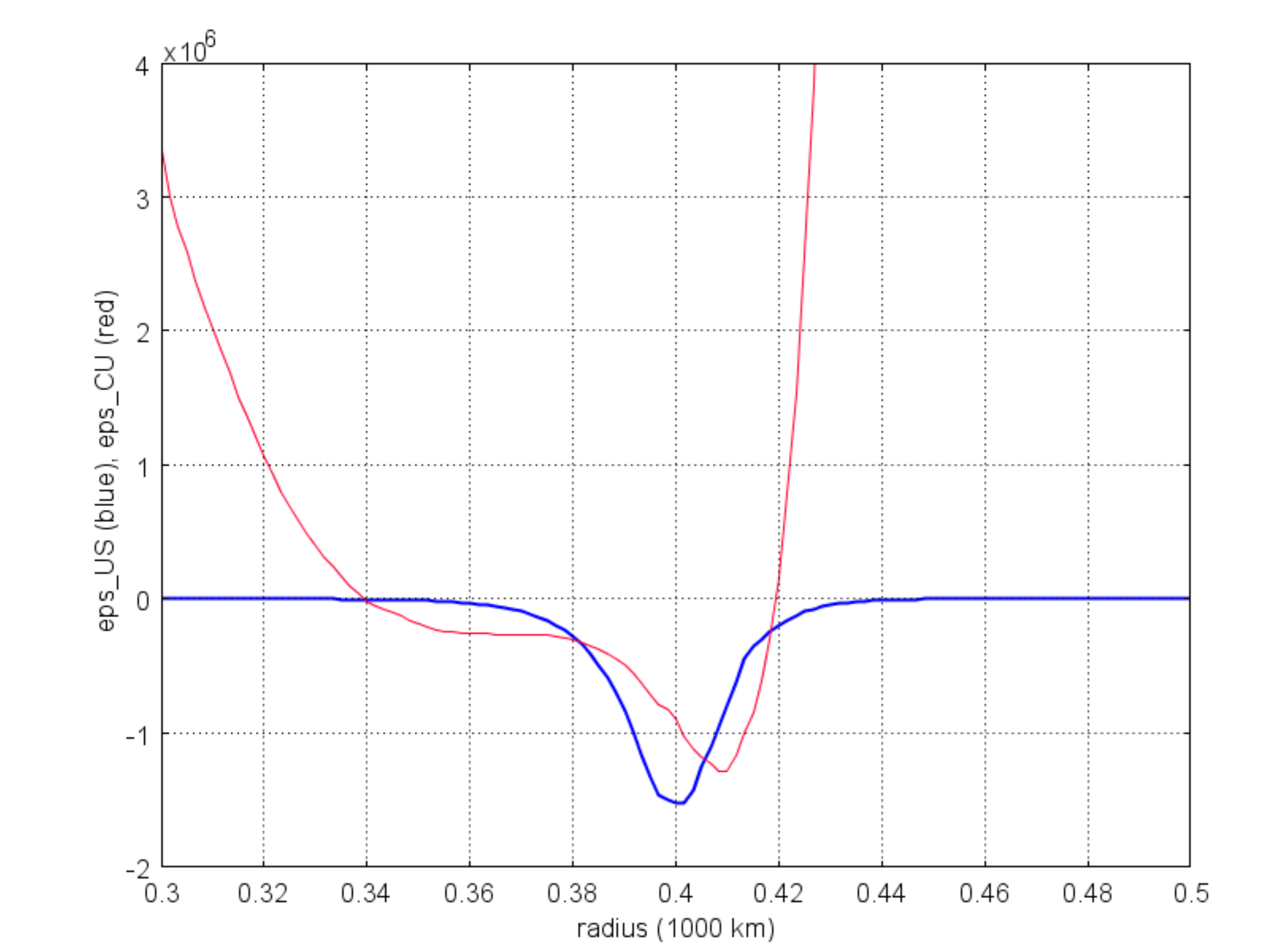}
\caption{\small The Urca cooling caused by neutrino and antineutrino escaping from the Urca shell (the blue curve)
        and the combined Urca cooling and heating (the red curve), the latter caused by non-equilibrium electron captures
        and beta decays around the Urca shell in the presence of convective mixing in the CO WD with $M_\mathrm{i} = 6.3\,M_\odot$.
        Both quantities are given in erg\,cm$^{-3}$\,s$^{-1}$.
        }
\label{fig:epsu_co_wd}
\end{figure}

Introducing the total number density of the Urca pair nuclei $n_\mathrm{U} = n_\mathrm{M} + n_\mathrm{D}$,
equation (\ref{eq:epsU}) can be rewritten in the following form:
\bea
\varepsilon_\mathrm{U} = C n_\mathrm{U} + H (n_\mathrm{M} - n_\mathrm{M}^{*}),
\label{eq:epsUCH}
\eea
where
\bea
n_\mathrm{M}^{*} = \frac{\lambda^-}{\lambda^+ + \lambda^-}\,n_\mathrm{U}
\eea
is an equilibrium number density of the mother nuclei, while
\bea
C = -\frac{L^+\lambda^- + L^-\lambda^+}{\lambda^+ + \lambda^-}
\eea
and
\bea
H = \delta\mu\,(\lambda^+ + \lambda^-) - L^+ + L^-
\eea
are the cooling and heating factors \citep{lesaffre:05}. Equation (\ref{eq:epsUCH}) shows that the Urca process
can cool the star down only in the vicinity of the Urca shell, where the first term is negative and the second term is close to
zero.  However, if the Urca shell is involved in convective mixing that redistributes the Urca nuclei accross the Urca shell, 
then the second term turns out to be always positive 
and it starts to dominate over the cooling term at some distance from the Urca shell (Fig.~\ref{fig:epsu_co_wd}). 
Therefore, such a ``convective Urca 
process'' should contribute to both cooling near the Urca shell and heating outside it \citep{bruenn:73}.

Although MESA does take into account the Urca energy loss/generation rate, 
e.g. as given by equation (\ref{eq:epsUCH}), 
it does not modify the MLT equations correspondingly, ignoring the fact that in the convective Urca process
the kinetic energy of convective motion serves as a source for both Urca cooling and heating \citep{bk:01}.
This means that in MESA the Urca process can affect convection only through changes of star's thermal structure.

Since the first discussion of convective Urca process by \cite{paczynski:72}, there has been found no satisfactory solution of 
this problem. The proposed solutions contradict each other, because some of them predict that the convective Urca process
should stabilize C burning, thus allowing the accreting WD to avoid the SN Ia explosion 
\citep{paczynski:72,iben:82,barkat:90}, while the others claim that it should rather destabilize
C burning sooner or later, thus leading to an earlier or delayed thermonuclear runaway 
\citep{bruenn:73,couch:75,mochkovitch:96,stein:99}.

The convective Urca process should probably decrease the volume occupied by convection that is driven by C burning during
the pre-explosion simmering phase \citep{lesaffre:05,stein:06}. However, we believe that a final solution of
the convective Urca problem and, in particular, a reliable estimate of the limit imposed by it on the size of the C convective
core can be obtained only in reactive-convective 3D hydrodynamical simulations. For example, these kinds of simulations of H
entrainment by He-shell flash convection in the post-AGB star Sakurai's object has recently revealed a very interesting and
complex mixing behaviour which could never be predicted by a mixing length theory \citep{herwig:14}. 
Until such simulations are done for the convective Urca process,
we choose to present the results of our calculations of SN Ia progenitor models for a number of different mixing assumptions 
that are supposed to mimic potentially possible outcomes of the interaction of convection and Urca reactions. 

\section{Results}
\label{sec:results}

We present the results only for three of our accreting WD models: the first one is the pure CO WD with the initial mass
$M_\mathrm{i} = 6.3\,M_\odot$ (the dashed black curve in Fig.~\ref{fig:hy_wd_c12}), 
the second is the hybrid WD with $M_\mathrm{i} = 7.3\,M_\odot$ that has a very small CO core 
left at the end of C burning (the red curve in Fig.~\ref{fig:hy_wd_c12}), 
and the third is the hybrid WD with $M_\mathrm{i} = 6.9\,M_\odot$ that has a relatively large CO core
(the blue curve in Fig.~\ref{fig:hy_wd_c12}).
The first model is a standard one for the SD channel of SN Ia progenitors. We include it for the completeness of our study.
The second model is close to the pure ONe WDs, therefore we expect that it will end up as an electron-capture
supernova (ECSN) \citep{jones:13}. The abundance profiles of the Urca mother nuclei $^{23}$Na and $^{25}$Mg for these models
are plotted in Fig.~\ref{fig:x23x25}. 

\subsection{The pure CO WD model with $M_\mathrm{i} = 6.3\,M_\odot$}

\begin{figure}
\includegraphics[width=9cm]{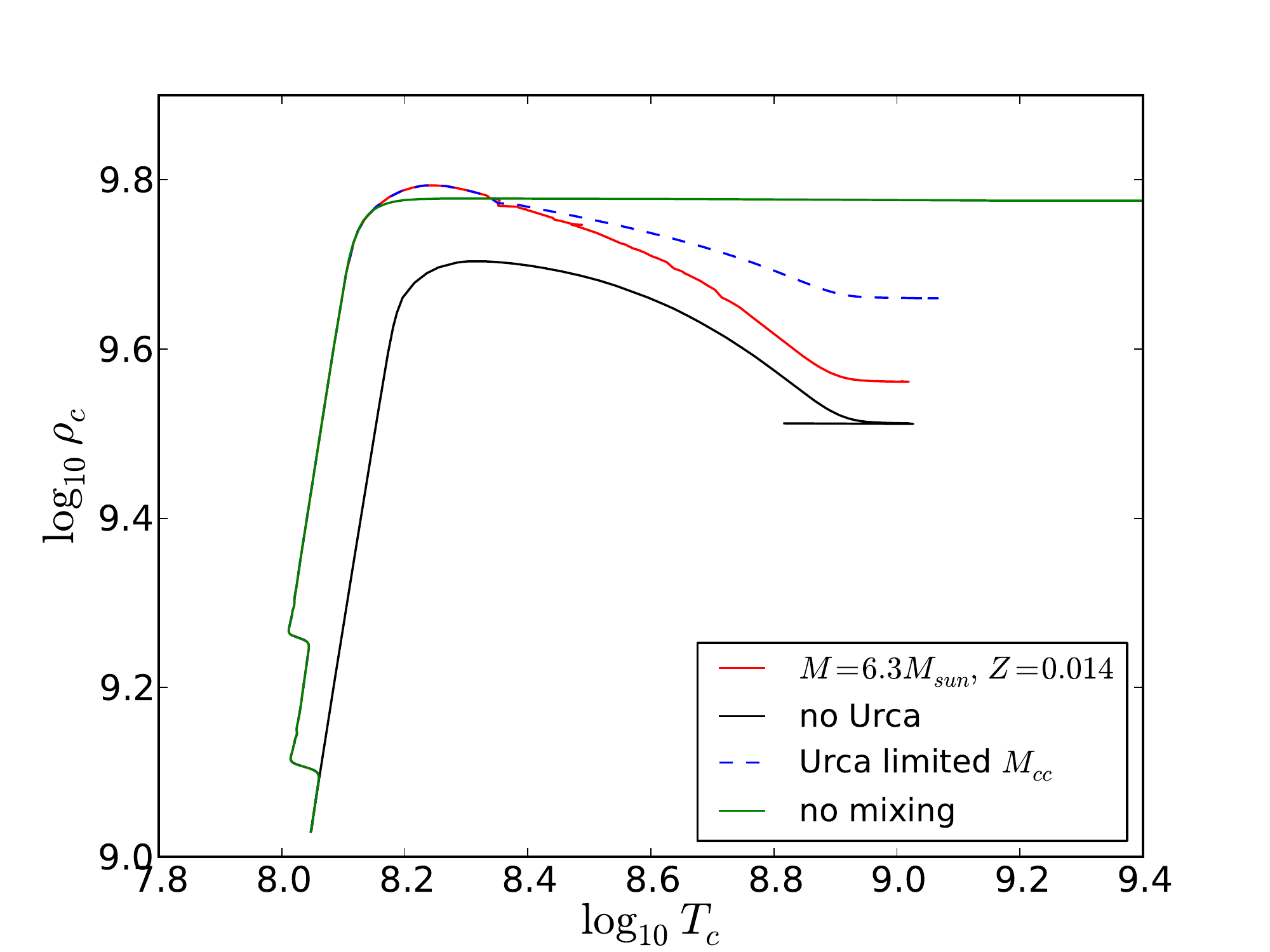}
\caption{\small Red curve shows the evolution of central temperature and density resulting from the accretion of C-rich
         surface-composition material onto the pure CO white dwarf with the rate  of $10^{-8}\,M_\odot\,{\rm yr}^{-1}$.
         The initial mass of the star is $6.3\,M_\odot$ (the dashed black curve in Fig.~\ref{fig:hy_wd_c12}).
         Other calculations did not include the Urca reactions (``no Urca''), assumed that the boundary of 
         the C-burning convective core $M_{\rm cc}$  is located at the $^{23}$Na/$^{23}$Ne Urca shell
         (``Urca limited $M_\mathrm{cc}$), or completely turned off convective mixing (``no mixing'').
         In the third and fourth cases, the density of explosive C ignition is higher than in the first two cases.
         }
\label{fig:TcRhoc_m63}
\end{figure}

The evolution of this WD model towards the explosive C ignition has been calculated relatively quickly 
with the mass accretion rate $\dot{M}_\mathrm{acc} = 10^{-8}\,M_\odot$\,yr$^{-1}$.
It is an order of magnitude lower than the corresponding value of $\dot{M}_\mathrm{stable}$
\citep{hachisu:96,nomoto:07,ma:13}, but, as we explained before,
a choice of evolutionary motivated values of $\dot{M}_\mathrm{acc}$ and $T_\mathrm{c}$ is out of the scope of the present study.
In spite of this, our calculated central temperature and density at the explosive C ignition near the right end of
the black curve in Fig.~\ref{fig:TcRhoc_m63} are very close to their corresponding values
$\log_{10}T_\mathrm{c} = 9.0$ and $\log_{10}\rho_\mathrm{c} = 9.4$ reported by \cite{han:14}, who chose the initial model
parameters more carefully. Our final WD mass $M_\mathrm{WD} = 1.386\,M_\odot$ is also close to their value of $1.387\,M_\odot$.
We have compared their final WD parameters with those obtained by us for the case without Urca processes (``no Urca'') because
\cite{han:14} did not include any Urca reactions.

The red curve in Fig.~\ref{fig:TcRhoc_m63} shows the evolution of $T_\mathrm{c}$ and $\rho_\mathrm{c}$ for our standard case, when
the Urca reactions are included using the new well-resolved tables of \cite{toki:13}. The two steps near the left ends of
all the curves that include the Urca processes are produced by sudden cooling at the centre happening when the increasing central 
density first exceeds the Urca threshold density for $^{25}$Mg and then the one for $^{23}$Na. Note that the red and black curves
would be indistinguishable if we used the old Urca reaction data of \cite{oda:94}.

To model the convective Urca process, \cite{lesaffre:05} developed a new two-stream formalism. For stationary conditions, they
found strong reductions of the convective velocity by the Urca process when the mass fraction of $^{23}$Na was increasing
from $X_{23} = 10^{-9}$ to $X_{23} = 10^{-3}$. Similarly, in their 2D hydrodynamical simulations of the convective Urca process, 
that they called
``an indicative but still qualitative study'' because of artificially increased reaction rates, 
\cite{stein:06} observed a confinement of the C convective core by the $^{23}$Na/$^{23}$Ne Urca shell 
for $X_{23} = 4\times 10^{-4}$. Taking these results into account and also the fact that our hybrid WD models
have very large abundances of $^{23}$Na (Fig.~\ref{fig:x23x25}), we consider as potentially possible outcomes of
the convective Urca process cases in which it either limits the mass of the C convective core $M_\mathrm{cc}$ by
a mass coordinate where the $^{23}$Na/$^{23}$Ne Urca shell is located or completely suppresses convective mixing.
For our mixing assumptions corresponding to these two cases that we call ``Urca limited $M_\mathrm{cc}$'' and ``no mixing'',
the resulting $T_\mathrm{c}$\,-\,$\rho_\mathrm{c}$ trajectories are also plotted in Fig.~\ref{fig:TcRhoc_m63}.

\begin{figure}
\includegraphics[width=9cm]{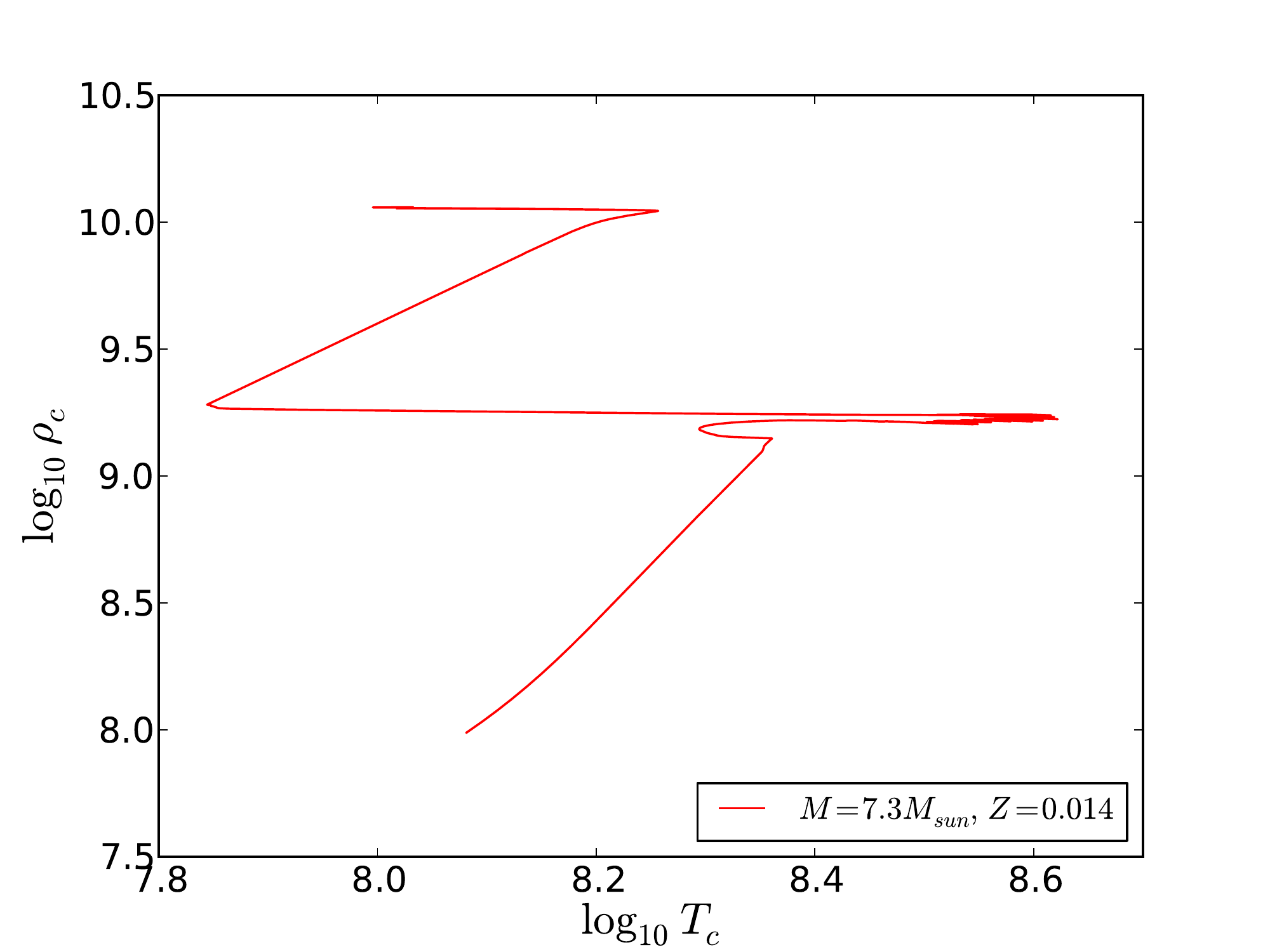}
\caption{\small Same as in Fig.~\ref{fig:TcRhoc_m63}, but for the hybrid C-O-Ne white dwarf with $M_\mathrm{i} = 7.3\,M_\odot$
        (the solid red curve in Fig.~\ref{fig:hy_wd_c12}) and for the mass accretion rate of
        $8\times 10^{-7}\,M_\odot\,{\rm yr}^{-1}$.  In this case, C fails to ignite explosively because of     
        its low abundance, and the star evolves towards electron-capture induced collapse.
        }
\label{fig:m73tc}
\end{figure}

\subsection{The hybrid WD model with $M_\mathrm{i} = 7.3\,M_\odot$ that has a small CO core}

When C is first ignited in the centre of this model, convection driven by the C burning dilutes the high C abundance 
in the small core with the low C abundance in the surrounding ONe zone (the red curve in Fig.~\ref{fig:hy_wd_c12}). 
The diluted C abundance $X_{12}\approx 0.02$ turns out to be too low
for the C burning to compete with the neutrino cooling and, as a result, the central temperature rapidly decreases
after its initial attempt to go up caused by the C ignition (the first zigzag at densities between 
$\log_{10}\rho_\mathrm{c} = 9$  and $\log_{10}\rho_\mathrm{c} = 9.5$ in Fig.~\ref{fig:m73tc}).
The continuing mass accretion with the rate $\dot{M}_\mathrm{acc} = 8\times 10^{-7}\,M_\odot$\,yr$^{-1}$, which is close
to the upper limit $\dot{M}_\mathrm{RG}$, unless we account for the thick-wind and super-Eddington regimes \citep{hachisu:96,ma:13},
leads to a further increase of the central density until it reaches a value 
at which electron-capture reactions become efficient (the second zigzag at $\log_{10}\rho_\mathrm{c} > 10$ in Fig.~\ref{fig:m73tc}).
Eventually, these reactions should induce a collapse of the star resulting in
an ECSN, like in the cases of more massive pure ONe cores of super-AGB stars considered
by \cite{jones:13}.

\subsection{The hybrid WD model with $M_\mathrm{i} = 6.9\,M_\odot$ that has a relatively large CO core}

This WD model is quite different from the previous two models. Indeed, on the one hand, it has a relatively large CO core
(the blue curve in Fig.~\ref{fig:hy_wd_c12}), so that
even after the dilution of C from the core in the ONe zone its abundance may still be high enough for the C burning to compete 
with the neutrino cooling. On the other hand,
$^{25}$Mg and, especially, $^{23}$Na have so high abundances outside the CO core in this model 
(the dashed and solid blue curves in Fig.~\ref{fig:x23x25}) that it is difficult to make   
a prediction, based on our 1D stellar evolution calculations with the MLT prescription for convection,
of how the convective Urca process will affect the model thermal and chemical structures.
Therefore, until appropriate reactive-convective 3D hydrodynamical simulations of 
this process provide more realistic prescriptions for the modeling of its effects on the explosive C ignition 
in the CO and hybrid C-O-Ne WDs in 1D stellar evolution codes, we have nothing else but to consider a number of instances of 
this accreting WD model obtained for different assumptions about possible outcomes of the interaction between convective mixing
and Urca reactions that might be suggested by such simulations in the future. 

\begin{figure}
\includegraphics[width=9cm]{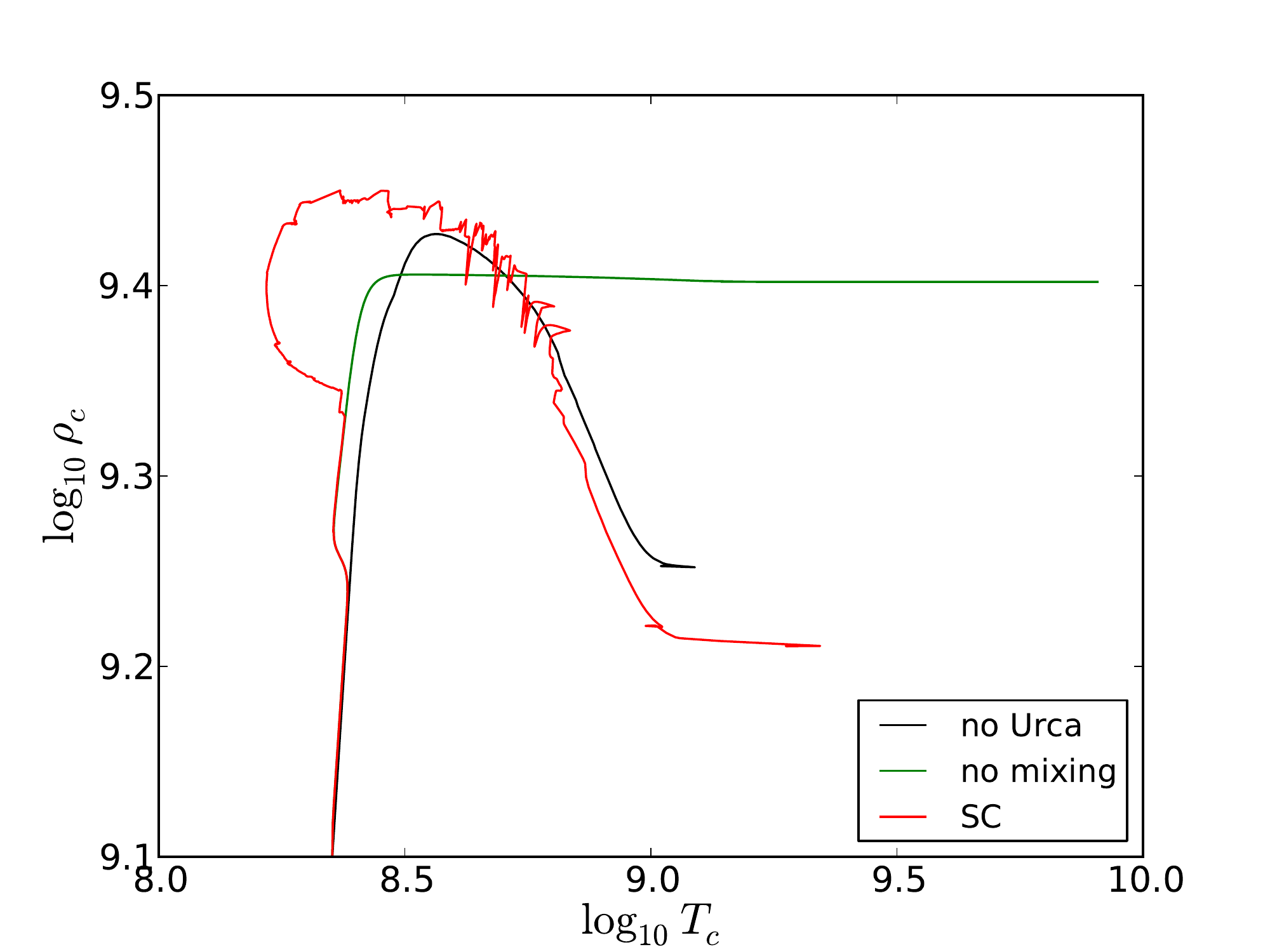}
\caption{\small Same as in Fig.~\ref{fig:TcRhoc_m63}, but for the hybrid C-O-Ne white dwarf with $M_\mathrm{i} = 6.9\,M_\odot$
        (the solid blue curve in Fig.~\ref{fig:hy_wd_c12}) and for the mass accretion rate of
        $8\times 10^{-7}\,M_\odot\,{\rm yr}^{-1}$.  In these three case, the explosive C ignition takes place in the centre.
        }
\label{fig:m69tc}
\end{figure}

\begin{figure*}
\includegraphics[bb = 100 140 500 700]{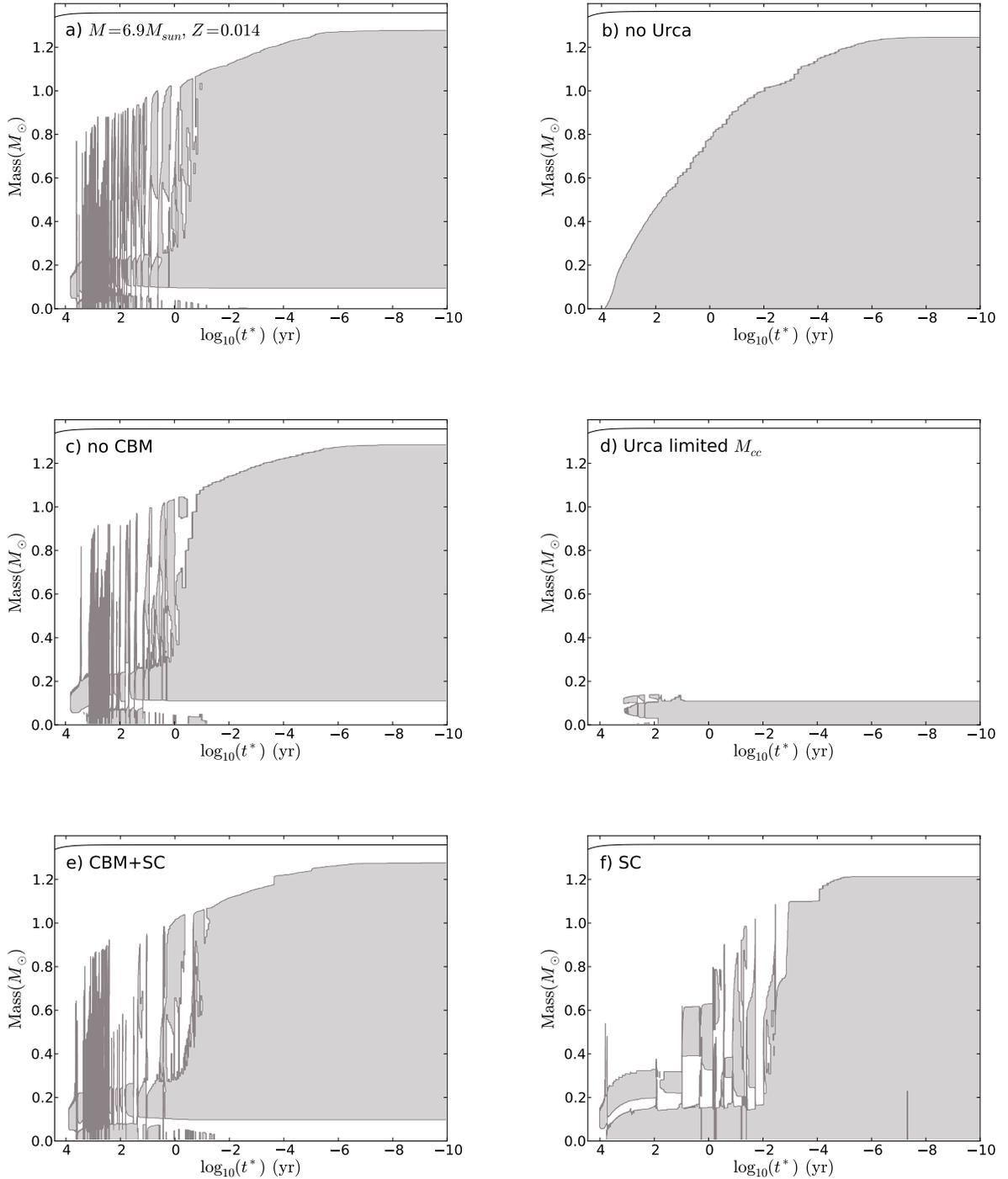}
\caption{\small Kippenhahn diagrams for the hybrid C-O-Ne white dwarf with $M_\mathrm{i} = 6.9\,M_\odot$
         (the solid blue curve in Fig.~\ref{fig:hy_wd_c12}) accreting C-rich material with the rate
        $8\times 10^{-7}\,M_\odot\,{\rm yr}^{-1}$ when it approaches the phase of the explosive C ignition.
        The diagrams are shown for the different mixing assumptions that are discussed in the text.
        The grey shades show convective zones.
         }
\label{fig:kipp}
\end{figure*}

Like the previous hybrid model, this WD model accretes C-rich material with the rate 
$\dot{M}_\mathrm{acc} = 8\times 10^{-7}\,M_\odot$\,yr$^{-1}$. We start with the case without Urca reactions.
Its corresponding WD's evolution on the $T_\mathrm{c}$\,-\,$\rho_\mathrm{c}$ plane is shown with the black curve in
Fig.~\ref{fig:m69tc}, while its Kippenhahn diagram is plotted in Fig.~\ref{fig:kipp}b. Similar to Fig.~\ref{fig:TcRhoc_m63},
the green curve in Fig.~\ref{fig:m69tc} has been calculated for the case when all mixing is assumed to be suppressed 
by the Urca processes.

\begin{figure}
\includegraphics[width=9cm]{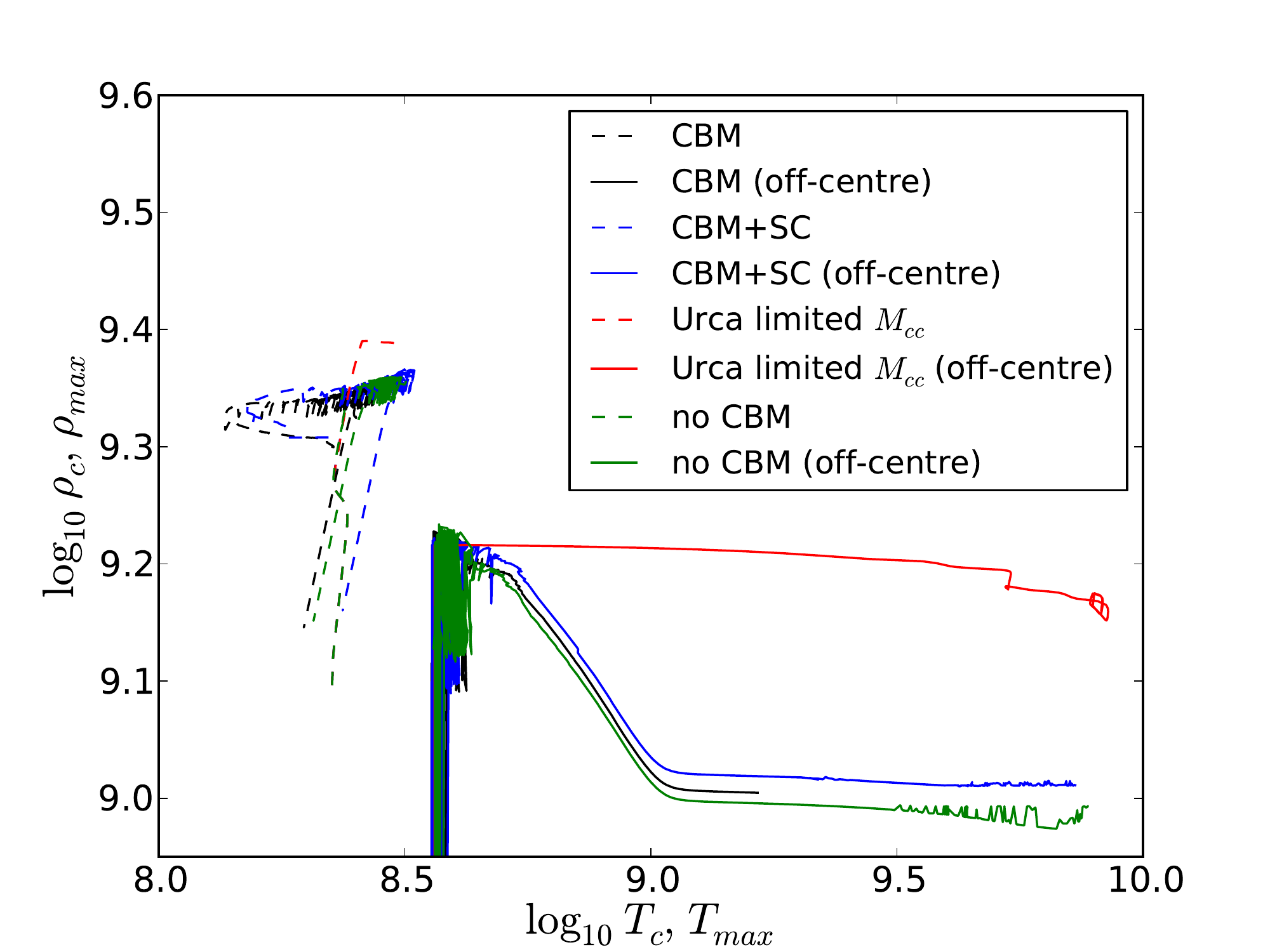}
\caption{\small Same as in Fig.~\ref{fig:TcRhoc_m63}, but for the hybrid C-O-Ne white dwarf with $M_\mathrm{i} = 6.9\,M_\odot$
        (the solid blue curve in Fig.~\ref{fig:hy_wd_c12}) and for the mass accretion rate of
        $8\times 10^{-7}\,M_\odot\,{\rm yr}^{-1}$.  In these cases, the explosive C ignition occurs off the centre as
        a result of Urca cooling in the centre, therefore two sets of curves are shown, one for the centre (dashed
        curves) and the other for the point where the temperature has a maximum (solid curves).
        }
\label{fig:m69tm}
\end{figure}

\begin{figure}
\includegraphics[width=9cm]{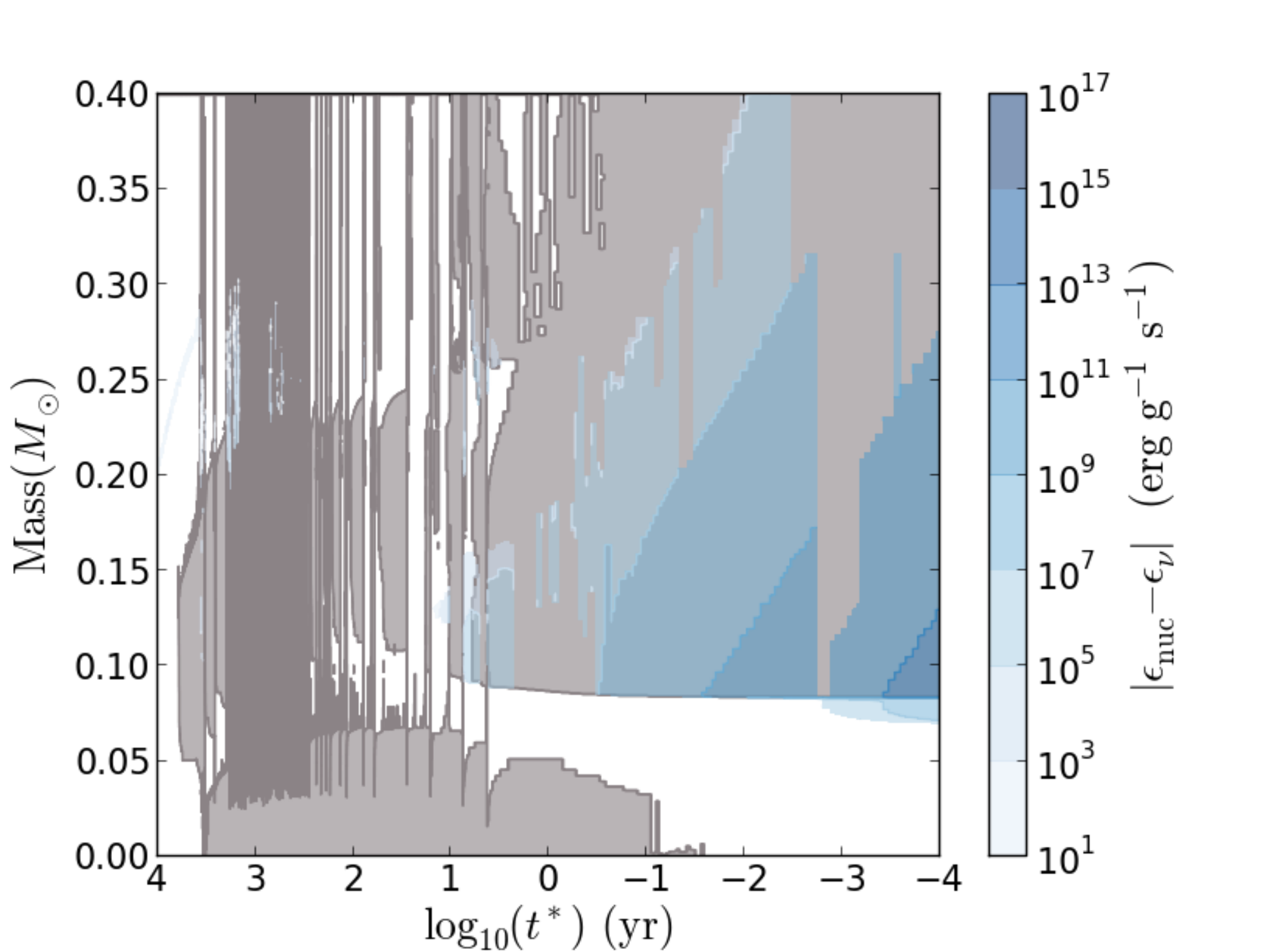}
\caption{\small Zoomed in Kippenhahn diagram from Fig.~\ref{fig:kipp}a. The added shades of blue show
         the energy generation by the Urca reactions and C burning.
         }
\label{fig:zoom}
\end{figure}

Our standard case with CBM produces a set of two black curves on the $T_\mathrm{c}$\,-\,$\rho_\mathrm{c}$ plane
in Fig.~\ref{fig:m69tm}. The Urca process involving the $^{23}$Na/$^{23}$Ne pair turns out to generate so efficient 
cooling in the centre of this hybrid WD model, when its mass approaches the Chandrasekhar limit, that the maximum temperature
is reached and the explosive C ignition occurs off the centre, as shown with the solid black curve in Fig.~\ref{fig:m69tm} and 
in Fig.~\ref{fig:kipp}a. This shift of $T_\mathrm{max}$ from the centre to a point where $\rho < \rho_\mathrm{th}$ for
the $^{23}$Na/$^{23}$Ne Urca pair and where the Urca process can therefore not interfere with the C burning is facilitated
by a series of convective Urca shell flashes near the centre. These flashes are seen in Fig.~\ref{fig:kipp}a and, especially,
in the zoomed in version of this figure in Fig.~\ref{fig:zoom}, which uses the shades of blue to additionally show
the positive energy generation rate. The convective shell flashes are caused by intermittent changes in the temperature 
profile taking place where the internal energy experiences local variations, as described by equation (\ref{eq:epsUCH}).
However, given that both Urca cooling and heating should probably drain convective motion of its kinetic energy
\citep{bk:01}, which is
not taken into account in our MLT prescription for convection, these ``convective Urca flashes'' may simply be an artefact of
our crude model. Their presence is not related to our using the CBM, because when we do not include the CBM they are still present
(the case of ``no CBM'' in Fig.~\ref{fig:m69tm} and Fig.~\ref{fig:kipp}c). The addition of semiconvection (the case
``CBM+SC''), following the prescription of \cite{langer:85} with the efficiency parameter $\alpha_\mathrm{sc} = 0.01$, 
does not change
qualitatively their behaviour. The semiconvection may be important near the boundaries of convective Urca shells, 
because Urca reactions build a gradient of the mean molecular weight that opposes a convectively unstable temperature gradient 
\cite[e.g.][]{lesaffre:05}, although
the CBM should dominate over semiconvection in regions where they are both present. Surprisingly, when we include only
semiconvection (``SC'') then $T_\mathrm{max}$ and the point of the explosive C ignition are not shifted from
the centre (Fig.~\ref{fig:kipp}f), however we believe that the CBM should always be included as well. Finally, in the case of 
``Urca limited $M_\mathrm{cc}$'' we again find an explosive C ignition off the centre (Fig.~\ref{fig:kipp}d and the solid
red curve in Fig.~\ref{fig:m69tm}).

\section{Discussion and conclusion}

\cite{denissenkov:13} and \cite{chen:14} have proposed a new class of WD models, called ``the hybrid C-O-Ne WDs'' that
consist of a CO core surrounded by a thick ONe zone. At their birth, they also have a thin CO-rich buffer zone at the surface.
Our stellar evolution calculations have shown that, when we include the observationally constrained extent of CBM outside
the H and He convective cores, as modeled by equation (\ref{eq:DCBM}) with $f = 0.014$, and assume that $f = 0.007$
outside the stiffer boundaries of the C convective zone in the core of a super-AGB star, the range of the initial mass of
the stars with $Z=0.014$ and $X=0.70$ that give birth to the hybrid WDs is between $M_\mathrm{i}\approx 6.4\,M_\odot$ and
$M_\mathrm{i}\approx 7.3\,M_\odot$ (Fig.~\ref{fig:hy_wd_c12}). The aim of this work has been to find out if the hybrid WDs
can be progenitors of SNe Ia, like their pure CO counterparts, provided that they are formed in binary systems with
parameters suitable for the SD channel. To answer this question, we have allowed two of our hybrid WD models, those with
$M_\mathrm{i} = 6.9\,M_\odot$ and $M_\mathrm{i} = 7.3\,M_\odot$, to accrete material with chemical composition
identical to that of their surface layers until C is ignited explosively or it fails to do so.
For the completeness of our study and for a comparison, we have also considered the evolution of the accreting CO WD 
with $M_\mathrm{i} = 6.3\,M_\odot$.

We have found that the most massive of our hybrid WD models, the one with $M_\mathrm{i} = 7.3\,M_\odot$ and, because of
the similarity of their $^{12}$C mass fraction profiles (the red and green curves in Fig.~\ref{fig:hy_wd_c12}), also the model 
with $M_\mathrm{i} = 7.1\,M_\odot$, 
will probably experience an electron-capture induced collapse leading to an ECSN, when its mass approaches
the Chandrasekhar limit. This happens with its slightly more massive pure ONe counterparts \citep{jones:13}.
This outcome is not surprising, given the small mass of the CO core in this star.
The non-explosive C ignition in its centre leads to the development of a C convective core, in
which the high C abundance from the CO core is diluted with the low C abundance in the ONe zone. As a result, the diluted
C abundance turns out to be so low that the energy generated in the C burning is surpassed by the neutrino cooling and,
after its initial attempt to go up, the central temperature begins to decline rapidly. The continuing increase of WD's mass
and associated increase of its central density will eventually make electron-capture reactions efficient enough to induce 
a collapse of the star.

The hybrid WD model with $M_\mathrm{i} = 6.9\,M_\odot$ has quite a large CO core, 
so that even after the carbon from the CO core is mixed with what is left of it in the ONe zone the resulting C abundance 
still remains high enough for the C burning to successfully compete with the neutrino
cooling. As a result, we find the explosive C ignition in all the substances of this WD model that we have obtained making
different assumptions about mixing. The most difficult problem that we encounter in this model is the convective Urca process
uncertainty. It has been known since the 1970s, but it is emphasized in our hybrid WD models because they have very high
abundances of the Urca mother nuclei $^{25}$Mg and, especially, $^{23}$Na, the latter being one of the main products of C
burning. The stellar evolution code of MESA that we use does account for the change of thermal energy produced by the Urca
reactions because it includes terms similar to (\ref{eq:epsUCH}) in its energy equation. However, the MLTs of convection used 
in MESA do not take into account the extraction of kinetic energy from convective motion by the Urca processes, which
is the major source of the Urca process uncertainty. Guided by the results of the 2D hydrodynamical simulations of 
\cite{stein:06} as well as by those obtained with the two-stream formalism by \cite{lesaffre:05}, and in
the absence of reactive-convective 3D hydrodynamical simulations of the convective Urca process, we have chosen to consider
a number of alternative instances of our accreting hybrid WD model for different mixing assumptions that are made to mimic the
potentially possible outcomes of the interaction between convection and Urca reactions. These assumptions include the two limiting
cases in one of which there are no Urca reactions at all (``no Urca''), while in the other Urca reactions completely suppress 
any mixing (``no mixing''). The mixing assumption suggested by the simulations of \cite{stein:06} is
the ``Urca limited $M_\mathrm{cc}$''. It limits the size of the C convective core by the location of the $^{23}$Na/$^{23}$Ne
Urca shell. We have also added the semiconvection. The $T_\mathrm{c}$\,-\,$\rho_\mathrm{c}$ trajectories of the four model
instances in which the explosive C ignition occurs off the centre are shown in Fig.~\ref{fig:m69tm}, while 
the $T_\mathrm{c}$\,-\,$\rho_\mathrm{c}$ trajectories of the three other model instances in which C explodes in the centre
are shown in Fig.~\ref{fig:m69tc}. Their corresponding Kippenhahn diagrams are gathered in Fig.~\ref{fig:kipp}.

From the presented results, we have come to the conclusion that the hybrid C-O-Ne WDs can be progenitors of SNe Ia, provided that
the initial masses of their parent stars are between $M_\mathrm{i}\approx 6.5\,M_\odot$ and $M_\mathrm{i}\approx 6.9\,M_\odot$.
However, their exact thermal and chemical structures at the moment of C explosion are uncertain and depend on the effects of
the Urca processes on convective mixing. These effects are most likely to be strongly non-linear \citep[e.g.][]{bk:01}, therefore 
we think that only future reactive-convective 3D hydrodynamical simulations will be able to estimate them quantitatively.
We have generated several potentially possible instances of the hybrid SN Ia progenitor model
that can be used, in the meanwhile, as initial models for 2D or 3D simulations of SN Ia explosion to see how 
their resulting nucleosynthesis
yields and light curves differ from each other and if they match any of available SN Ia observations.
The hybrid WDs should probably lead to SNe Ia that are different in some aspects from those produced from the pure CO WDs because
they have much lower C/O abundance ratios and a slightly smaller electron-to-nucleon ratio $Y_\mathrm{e}$, the latter being affected
by the high $^{23}$Na abundance \citep{timmes:03}. Accounting for their existence should also modify the predicted 
SN Ia and ECSN rates.

Although it can be expected that the less massive of our stars will enter a phase of explosive C-burning once their cores 
approach the Chandrasekhar mass, it is unlikely that the outcome will be a normal SN Ia. If the burning would proceed 
as a detonation from the start, the lack of intermediate-mass elements in the ejecta would, like in the pure CO white dwarfs, 
contradict observations. In contrast, a deflagration ignited in the CO core cannot change into a detonation
in the ONe zone easily because the critical mass for a detonation of an O and Ne mixture is much larger than that of carbon. 
Therefore, the more likely outcome of such an explosion is a faint SN Ia, similar to the SN 2002cx class, as was also found 
for pure-deflagration CO white dwarfs, and possibly a bound remnant is left behind \cite[][and references therein]{fink:14}.

Estimates of SNIa birthrates with our hybrid WD
progenitors have recently been made by \cite{meng:14} and \cite{wang:14}.
They have found that these rates can range from a few to 18\%, depending on the assumptions about the carbon burning rate,
which is still very uncertain \citep{chen:14}, chemical composition of the accreted material and details of the used
binary population synthesis model.

\section*{Acknowledgments} 

This research has been supported by the National
Science Foundation under grants PHY 11-25915 and AST 11-09174. This
project was also supported by JINA (NSF grant PHY 08-22648). 
Falk Herwig acknowledges funding from Natural Sciences and Engineering Research Council of
Canada (NSERC) through a Discovery
Grant. Pavel Denissenkov thanks Wolfgang Hillebrandt for his comments.
Ken Nomoto acknowledges the support from the Grant-in-Aid for
Scientific Research (23224004, 23540262, 26400222) from the Japan
Society for the Promotion of Science, and the World Premier
International Research Center Initiative (WPI Initiative), MEXT, Japan.

\bibliography{paper}

\label{lastpage}

\end{document}